\documentclass{kluwer}    % Specifies the document style.
\usepackage{epsfig}
\newdisplay{guess}{Conjecture}
\newcommand{\etal} {{\it et al.}\ }

\begin{document}                                                                                   
\begin{article}
\begin{opening}         
\title{Population Synthesis and the Diagnostics of High-redshift Galaxies} 
\author{Alberto \surname{Buzzoni}}  
\runningauthor{A. Buzzoni}
\runningtitle{Diagnostics of high-redshift galaxies}
\institute{Telescopio Nazionale Galileo, A.P. 565, 38700 S/Cruz de La Palma (Tf), Spain and\\
Osservatorio Astronomico di Brera, Milano Italy}
\date{ }

\begin{abstract}
The effect of redshift on the observation of distant galaxies is briefly discussed emphasizing 
the possible sources of bias in the interpretation of high-$z$ data. A general energetic criterion
to assess physical self-consistency of evolutionary population synthesis models is also proposed,
for a more appropriate use of this important tool to investigate distinctive properties of primeval galaxies.
\end{abstract}
\keywords{Galaxies: high-redshift, evolution, stellar content}
\end{opening}           

\section{Introduction}

The recent major advances in the observation of the deep Universe have substantially improved
our chance to reach the redshift of galaxy formation.
Objects at $z > 5$ begin in fact to emerge from HST observations and other deep surveys with
the new-generation telescopes (Cohen \etal 2000).

As far as we move to larger distances, however, optical (and infrared) observations probe galaxy 
spectral energy distribution (SED) at much shorter restframe wavelength, in the ultraviolet domain. 
If neglected, this change in our spectral vantage point could lead to an important bias
in the interpretation of high-redshift data. In this note, we will briefly analyze some
aspects of this problem dealing both with the observational side and the appropriate
use of theoretical tools such as population synthesis models to match primeval galaxy evolution.

\section{Redshift constraints and galaxy spectral properties}

One important consequence of redshift (besides the first and most obvious effect of moving 
spectral features to longer wavelength) is that the wavelength interval $\Delta \lambda_{\it obs}$ 
sampled by any fixed set of photometric bands will span a narrower portion of the restframe SED when
observing galaxies at increasing $z$, such as $\Delta \lambda_{\it rest} = \Delta \lambda_{\it obs}/(1+z)$.

\begin{figure}
\epsfig{file=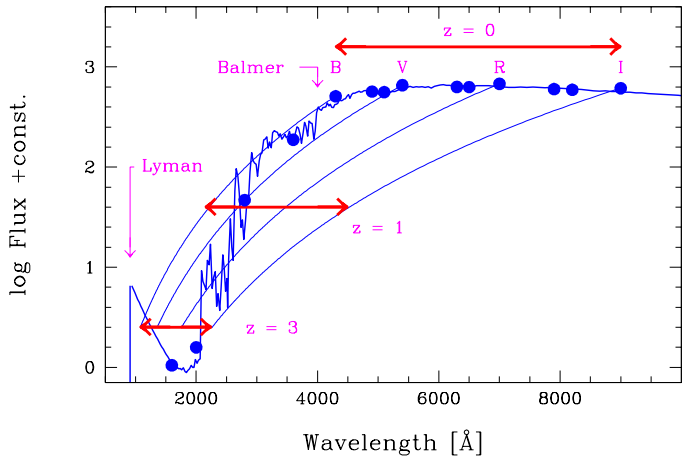,width=0.82\hsize,clip=}
\caption{Restframe portion of galaxy SED explored at different redshift by the Johnson four-color photometry
through the $B$, $V$, $R$, and $I$ bands. The wide wavelength range, between 4000 and 9000 \AA,
attained for local galaxies, ``shrinks'' to a 1000--2200 \AA\ interval when observing objects at $z = 3$.
The synthetic SED of a model elliptical is displayed, for reference.}
\label{f1}
\end{figure}

This effect, by itself, can be of paramount importance as far as we try to gain information
on galaxy stellar content by means of multicolor photometry. Figure~\ref{f1} gives an illuminating 
example in this sense. 

For a $z = 0$ galaxy, for instance, Johnson four-color photometry in the {\it BVRI} 
bands would span a wide wavelength range, from 4000 to 9000 \AA, and collect a substantial contribution 
from the galaxy stellar population. 
On the contrary,  when observing at $z = 3$ with the same photometric system, we
would be covering a scarce $\pm 600$~\AA\ spectral window centered about 1600 \AA\ in the galaxy
restframe. Quite deceivingly, our ``multicolor'' photometry at $z = 0$ would translate into a nearly 
``monochromatic'' estimate of the galaxy flux at $z = 3$.

\subsection{Sampling stellar populations of high-$z$ galaxies}

According to the overall color-magnitude distribution, each star in a galaxy aggregate
contributes with a different weight to the integrated luminosity with varying wavelength.
In Buzzoni (1993) we approached this problem in a statistical way introducing the concept of 
``effective'' contributors in a stellar population.
This quantity directly relates to galaxy integrated properties, such as total luminosity or
surface brightness fluctuations (as further explored, for instance, by Tonry and Schneider 1988).

We showed, in particular, that by sampling  at a given wavelength a total luminosity $L_{\it tot}$ provided 
by $N$ stars, the expected Poissonian fluctuation results:
\begin{equation}
{\sigma(L) \over L_{\it tot}} = {{(\sum^N l_i^2)^{1/2}}\over{\sum^N l_i}},
\end{equation}
where $l_i$ is the luminosity of each composing star.

\begin{figure}
\epsfig{file=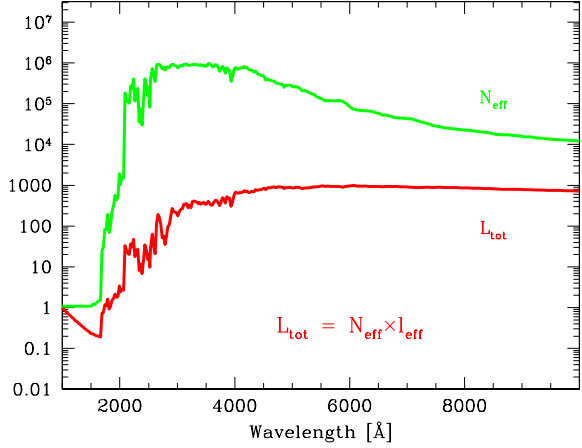,width=0.7\hsize,clip=}
\caption{Relative number of ``effective'' stellar contributors ($N_{\it eff}$)
and SED ($L_{\it tot}$) for a 5 Gyr SSP model with Salpeter IMF and [Fe/H]~$ = +0.3$ dex
according to Buzzoni (1993). Both quantities are in logarithm scale and arbitrarily normalized to 
their values at 1000 \AA.}
\label{f2}
\end{figure}

If all stars have the same luminosity, then simply $\sigma(L) / L_{\it tot} = 1/\sqrt{N}$.
More generally, however, a quantity 
\begin{equation}
N_{\it eff} = {{(\sum^N l_i)^2} \over {\sum^N l_i^2}} \leq N
\end{equation}
can be defined, as a function of $\lambda$. Its value gives a statistical measure of
the effective number of stars that provide $L_{\it tot}$ at a given wavelength.
Following Buzzoni (1993), a ``mean'' luminosity 
($l_{\it eff}$) for the composing stars can also be computed as
\begin{equation}
l_{\it eff} = {{\sum^N l_i^2}\over{\sum^N l_i}}
\end{equation}
so that, at every wavelength,
\begin{equation}
L_{\it tot} = N_{\it eff} \times l_{\it eff}.
\end{equation}

In Fig.~\ref{f2} we show the expected value of $N_{\it eff}$ for the illustrative
case of a 5 Gyr simple stellar population (SSP) with Salpeter IMF and [Fe/H]~$= +0.3$
along with the integrated SED of the model.
This should closely resemble the real case of primeval elliptical galaxies (Buzzoni 1995).

As a stricking feature of the model, note that below 2000 \AA\ the effective number of ``luminous'' contributors
dramatically drops by over six (!) orders of magnitude compared for example with the $B$ band value at 4000 \AA. 
On the contrary, luminosity over the same wavelength range only dims by a factor of $L_B/L_{2000} \sim 10^3$.
In terms of surface brightness fluctuations this means, for instance, that the ultraviolet image of a galaxy 
consisting of $10^{11}$ stars should fairly compare with the $B$ observations of a ($10^6$ star-poorer) 
Galactic globular cluster.

Such a severe undersampling of the galaxy stellar population in the ultraviolet range could have an important
impact when studying morphology of $z > 1$ ellipticals at optical wavelength (but the same argument holds also for
late-type systems). In particular, high-redshift objects would more likely appear as coarser systems with 
increasing $z$ (e.g.\ Puerari 2001, this conference).

\subsection{Redshift and age bias}

The selective sampling of galaxy stellar population with changing wavelength
has even more pervasive consequences, as far as we track evolution of starforming
systems at different redshift.

More generally, we could describe a composite stellar population in terms 
of a suitable convolution of SSPs according to the star formation rate (SFR) at the different epochs.
The total luminosity at age ``$t$'' results therefore:
\begin{equation}
L_{\it tot}(t) = \int_0^t L_{\it SSP}(\tau)\ {\it SFR}(t-\tau)\ d\tau
\end{equation}

A  mean luminosity-weighted age of the composing SSPs can also be derived such as
\begin{equation}
\overline{t}_* = {{\int_0^t \tau\ L_{\it SSP}(\tau) {\it SFR}(t-\tau)\ d\tau} \over {L_{\it tot}(t)}}
\label{eq:ave}
\end{equation}
This could be regarded as the ``representative'' age of the stars that contribute to galaxy
luminosity. We know (Buzzoni 1995; Tinsley and Gunn 1976) that SSP luminosity evolution can be pretty well 
described by a simple power law such as $L_{\it SSP} \propto t^{-\alpha}$ with the power index $\alpha$ that
is a function of wavelength. 
As an instructive example, if we consider for the SFR a power-law time decay such as 
SFR~$\propto t^{-\eta}$,\footnote{The major advantage of this parameterization is that 
the full range of galaxy morphologies can simply be accounted for by an
age-independent distinctive birthrate: $b = {\rm SFR}/<{\rm SFR}> = (1-\eta)$. This easily derives
recalling that, at age $t$, ${\rm SFR} = C\ t^{-\eta}$ and its time average is
$<{\rm SFR}> = C\ t^{-1}\int_0^t \tau^{-\eta}\ d\tau$.
Providing that $\eta < 1$, we always have $<{\rm SFR}> = {\rm SFR}/(1-\eta)$, from which the value
of $b$ directly follows, by definition.} eq.~(\ref{eq:ave}) becomes
\begin{equation}
\overline{t}_* = {{\int_0^t \tau^{1-\alpha}\ (t-\tau)^{-\eta}\ d\tau} \over {\int_0^t \tau^{-\alpha}\ (t-\tau)^{-\eta}\ d\tau}}.
\label{eq:ave2}
\end{equation}
The integral has an analytical solution such as
\begin{equation}
\overline{t}_* = {{1-\alpha}\over {2-\alpha-\eta}}\ t.
\label{eq:ave3}
\end{equation}
In bolometric, $\alpha = 0.76$ (Buzzoni 1989) and, for a constant SFR (i.e.\ $\eta = 0$),
$\overline {t}_* = 0.2\ t$. At 15 Gyr, bright stars are therefore in average 3 Gyr old, but this value changes 
according to the $\alpha(\lambda)$ function. For a SSP with Salpeter IMF and $Z_\odot$, the luminosity power
index smoothly increases from $\alpha = 0.71$ in the $K$ band to 0.94 in the $B$ band, and exceeds unity 
below 3500 \AA. A value of $\alpha \geq 1$ implies that $\overline{t}_* \to 0$, that is only the youngest and 
more massive stars are representative contributors to galaxy ultraviolet luminosity, as discussed
in the previous section. 

\begin{figure}
\epsfig{file=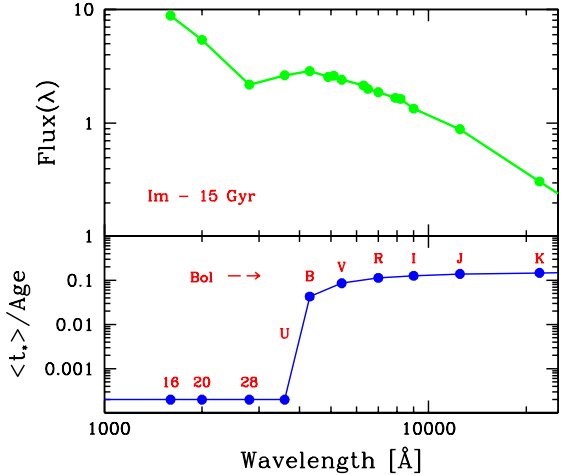,width=0.67\hsize,clip=}
\caption{Theoretical SED  ({\it upper panel}), and
``mean'' age (relative to galaxy age) of stars at different photometric bands (and in bolometric
as well), according to eq.~(\ref{eq:ave2}) ({\it lower panel}), for a 15 Gyr Im model galaxy from Buzzoni (2001).}
\label{f3}
\end{figure}

The mean age of stars contributing to galaxy luminosity at different wavelength for a synthesis model of a 
15 Gyr Magellanic irregular (see Buzzoni 2001 for details) is reported in Fig.~\ref{f3}. Note from the figure that
the value of $\overline{t}_*$ smoothly decreases with decreasing wavelength. As a result, high-redshift objects 
forcedly appear to be younger than local homologues at $z = 0$ {\it in spite of any intrinsic evolution.}

\section{Energetic self-consistency of population synthesis models}

Evolutionary population synthesis models are extensively used as a reference tool to reproduce galaxy colors 
and investigate galaxy spectrophotometric properties at the different cosmic epochs (Buzzoni 1989, 1995, 2001; 
Bruzual and Charlot 1993; Worthey 1994; Bressan \etal 1994; Fioc and Rocca-Volmerange 1997).

Besides any difference in the computational details and input physics,
a crucial constraint that needs to be consistently fulfilled in the synthesis procedure concerns a suitable
match of the MS and Post-MS luminosity contribution in the SSP models.
The so-called ``Fuel consumption theorem'' of Renzini and Buzzoni (1986) provides a simple and
very powerful tool in this sense.
According to the theorem, the Post-MS bolometric luminosity in a SSP of total luminosity
($L_{\it bol}$) is
\begin{equation}
L_{\it PMS} = {\cal B} \times L_{\it bol} \times {\rm Fuel}.
\label{eq:b}
\end{equation}
In the equation, the nuclear fuel refers to the amount consumed by one star leaving the MS turn-off (TO) point 
along the whole Post-MS evolution. This is a natural output of the theoretical stellar tracks
and is therefore known with high accuracy.
The scaling factor ${\cal B}$ is the ``specific evolutionary flux'' of the SSP, and directly deals with the
stellar clock, that is the relationship between MS lifetime and TO stellar mass (see Renzini and Buzzoni 1986).
In particular, we have that
\begin{equation}
{\cal B} \propto M_{TO}^{-s} \times \dot{M}_{TO}
\label{eq:b2}
\end{equation}
assuming that a power-law IMF holds with $dN_* \propto M_*^{-s}\ dM_*$ ($s = 2.35$ for the Salpeter case).
Again, the r.h.\ term of eq.~(\ref{eq:b2}) is a direct output of stellar evolution theory, and
$\cal B$ should therefore be regarded as a ``universal'' function, that is nearly insensitive to the specific
input physics of the SSP models.

\begin{figure}
\epsfig{file=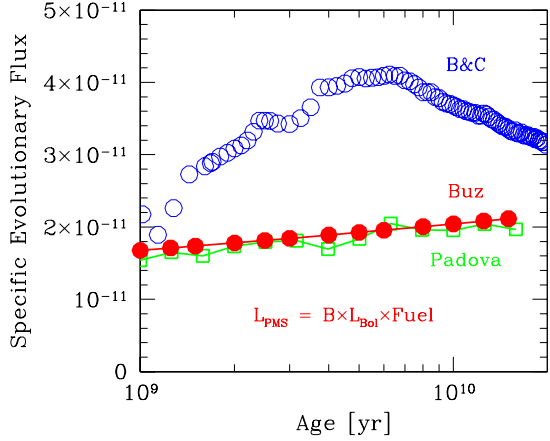,width=0.64\hsize,clip=}
\caption{The ``specific evolutionary flux'', $\cal B$, for SSP
models with $Z_\odot$ and Salpeter IMF according to Buzzoni (1989; ``$\bullet$'' markers), 
Bressan \etal (1994; ``$\square$'') and Bruzual and Charlot (1993; ``$\circ$''). 
The value of $\cal B$ is given in unit of yr$^{-1}$\ L$_\odot^{-1}$.
It directly scales Post-MS contribution to SSP luminosity through eq.~(\ref{eq:b}).}
\label{f4}
\end{figure}

\begin{figure}[b]
\epsfig{file=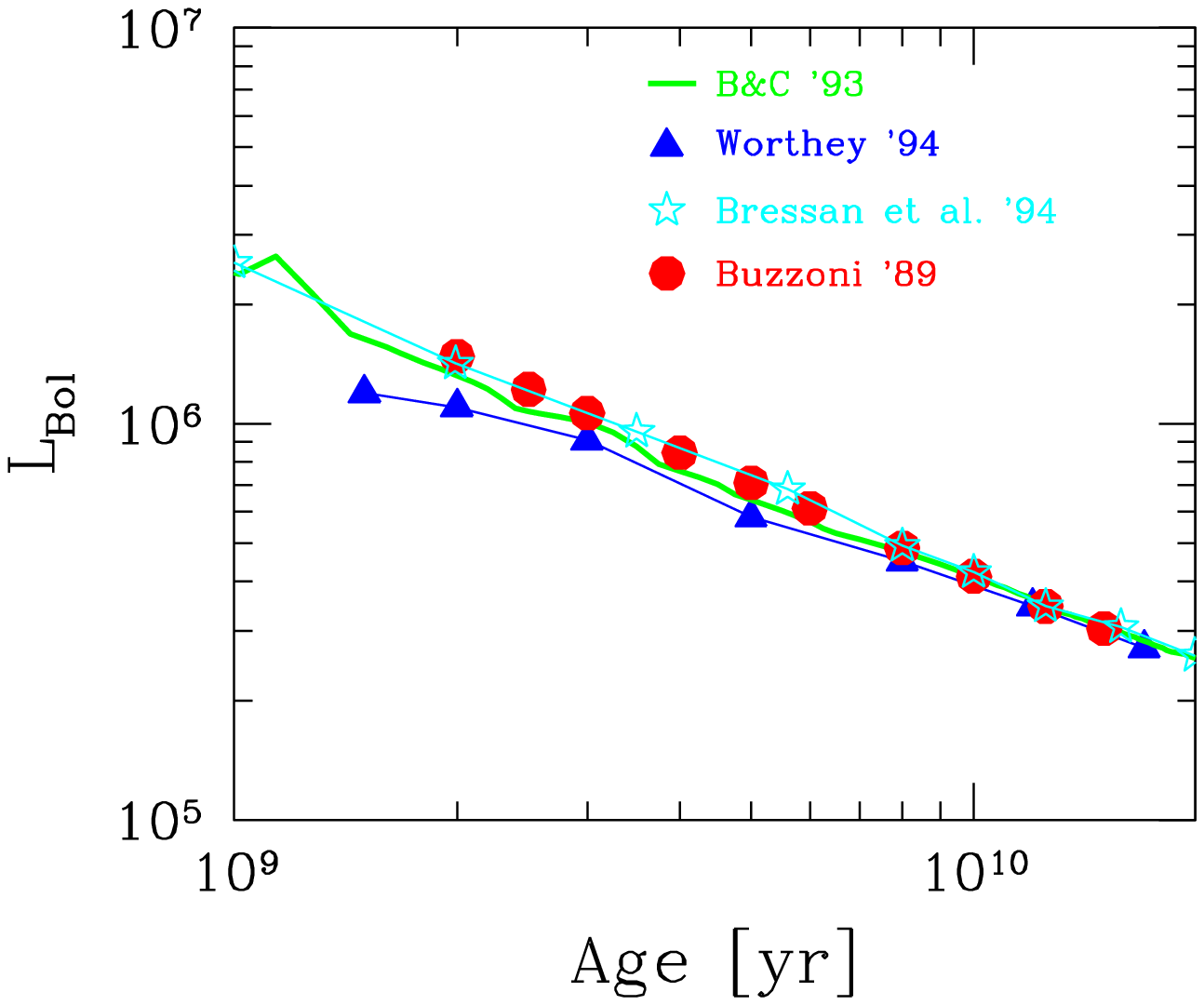,width=0.5\hsize,clip=}
\epsfig{file=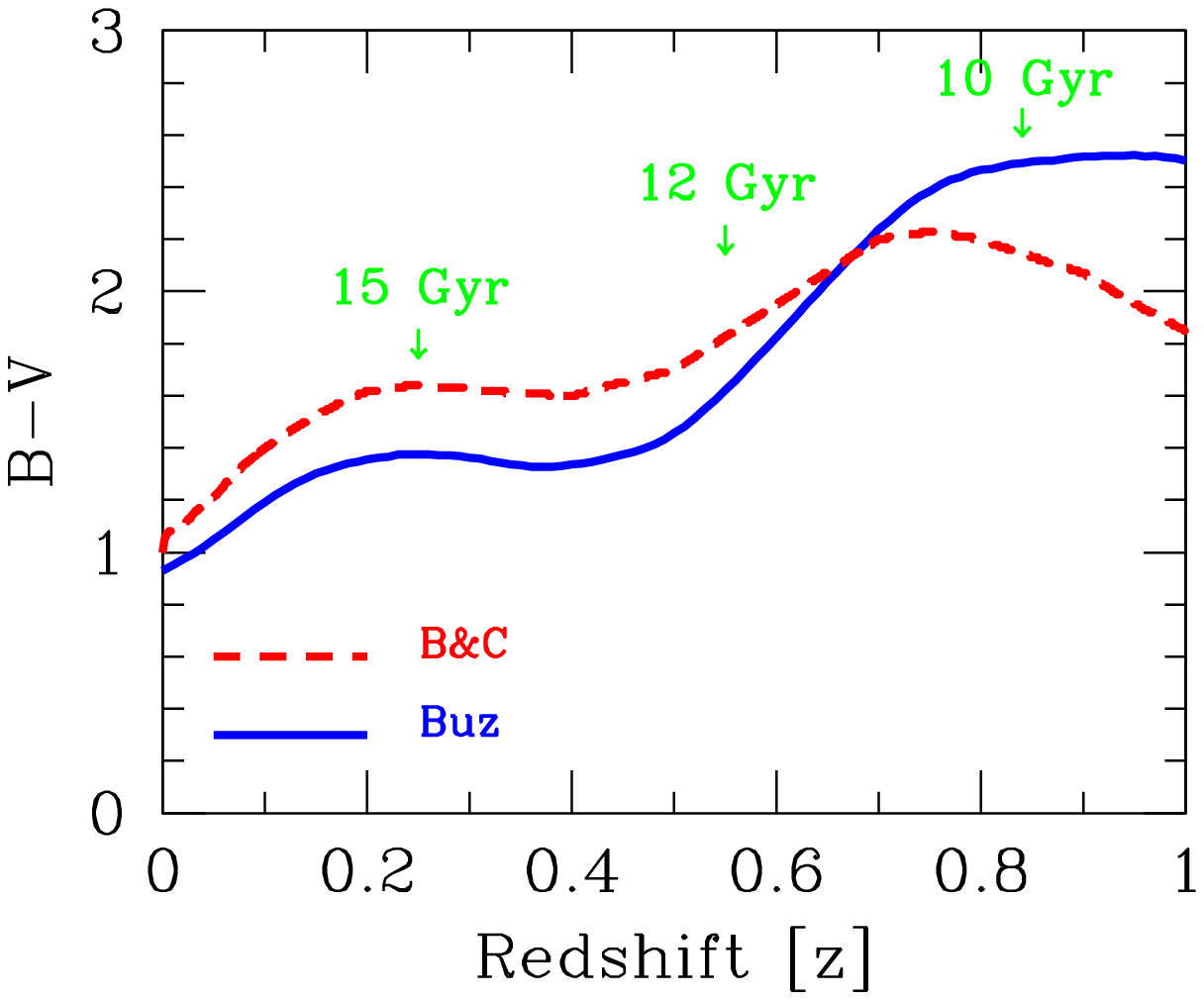,width=0.5\hsize,clip=}
\caption{{\it Left panel:} bolometric evolution for a SSP with $Z_\odot$ and Salpeter IMF according to different 
synthesis codes, as labelled. Luminosity is normalized at 12 Gyr.
{\it Right panel:} apparent $B-V$ evolution of elliptical galaxies according
to Buzzoni (1995; ``Buz'') and Bruzual and Charlot (1993; ``B\&C''). A cosmological model
with ($H_o, q_o, z_f$) = (50 km s$^{-1}$ Mpc$^{-1}$, 0, 30) is assumed.
The absolute age of galaxies at the different redshift is also labelled on the plot.}
\label{f5}
\end{figure}

A comparison of the $\cal B$ scaling factor, as derived from the synthesis codes of Bruzual and Charlot (1993),
Buzzoni (1989), and the Padova group (Bressan \etal 1994) for a SSP with $Z_\odot$ and Salpeter IMF
is displayed in Fig.~\ref{f4}. While both the Buzzoni and Padova codes consistently match
the fuel consumption theorem, the Bruzual and Charlot models show a much higher value for the specific
evolutionary flux. The effect is a macroscopic one, and reaches about a 100\%
discrepancy at intermediate age. Such higher scale factor leads, in eq.~(\ref{eq:b}), to a corresponding
overestimate of Post-MS luminosity at every age. The effect is quite shifty as it will not explicitly appear
when considering luminosity evolution (cf.\ Fig.~\ref{f5}, left panel), but would strongly affect 
the theoretical prediction of galaxy M/L ratio and apparent color evolution. In the latter case,
enhanced Post-MS contribution would lead to a larger number of red giant
stars and therefore a systematically redder color for the stellar aggregate as a whole.
This is shown in Fig.~\ref{f5} (right panel) when comparing the Bruzual and Charlot (1993) results with the 
expected evolution of elliptical galaxies according to Buzzoni (1995).

\acknowledgements
It is a pleasure to thank the INAOE for the kind invitation and the excellent organization of this meeting.
This work received partial financial support from the Italian MURST under COFIN '00 grant.

\end{article}
\end{document}